\documentclass[a4paper,clock]{article}
\usepackage[dvips]{epsfig}
\usepackage{amssymb}
\input{epsf}
\usepackage{bm}
\usepackage{dcolumn}

\catcode`\@=11
\def\marginnote#1{}
\newcount\hour
\newcount\minute
\newtoks\amorpm
\hour=\time\divide\hour by60 \minute=\time{\multiply\hour by60
\global\advance\minute by-\hour}\edef\standardtime{{\ifnum\hour<12
\global\amorpm={am}%
        \else\global\amorpm={pm}\advance\hour by-12 \fi
        \ifnum\hour=0 \hour=12 \fi
        \number\hour:\ifnum\minute<10
0\fi\number\minute\the\amorpm}}
\edef\militarytime{\number\hour:\ifnum\minute<10 0\fi\number\minute}

\def\draftlabel#1{{\@bsphack\if@filesw {\let\thepage\relax
   \xdef\@gtempa{\write\@auxout{\string
      \newlabel{#1}{{\@currentlabel}{\thepage}}}}}\@gtempa
   \if@nobreak \ifvmode\nobreak\fi\fi\fi\@esphack}
        \gdef\@eqnlabel{#1}}
\def\@eqnlabel{}
\def\@vacuum{}
\def\draftmarginnote#1{\marginpar{\raggedright\scriptsize\tt#1}}
\def\draft{\oddsidemargin -.5truein
        \def\@oddfoot{\sl preliminary draft \hfil
        \rm\thepage\hfil\sl\today\quad\militarytime}
        \let\@evenfoot\@oddfoot \overfullrule 3pt
        \let\label=\draftlabel
        \let\marginnote=\draftmarginnote

\def\@eqnnum{(\theequation)\rlap{\kern\marginparsep\tt\@eqnlabel}%
\global\let\@eqnlabel\@vacuum}  }

\def\numberbysection{\@addtoreset{equation}{section}
        \def\theequation{\thesection.\arabic{equation}}}

\def\underline#1{\relax\ifmmode\@@underline#1\else
 $\@@underline{\hbox{#1}}$\relax\fi}

\catcode`@=12 \relax

\numberbysection

\topmargin by -\headsep

\def\br{\begin{eqnarray}}
\def\er{\end{eqnarray}}
\def\be{\begin{equation}}
\def\ee{\end{equation}}

\def\({\left(}
\def\){\right)}

\relax

\def\tp0{\Theta_{+}^{(0)}}
\def\tm0{\Theta_{-}^{(0)}}

\def\f#1#2#3 {f^{#1#2}_{#3}}

\def\win1{{\sf w_{1+\infty}}}

\def\Win1{{\sf W_{1+\infty}}}

\def\rlx{\relax\leavevmode}
\def\inbar{\vrule height1.5ex width.4pt depth0pt}
\def\IZ{\rlx\hbox{\sf Z\kern-.4em Z}}
\def\IR{\rlx\hbox{\rm I\kern-.18em R}}
\def\IC{\rlx\hbox{\,$\inbar\kern-.3em{\rm C}$}}
\def\IN{\rlx\hbox{\rm I\kern-.18em N}}
\def\IO{\rlx\hbox{\,$\inbar\kern-.3em{\rm O}$}}
\def\IP{\rlx\hbox{\rm I\kern-.18em P}}
\def\IQ{\rlx\hbox{\,$\inbar\kern-.3em{\rm Q}$}}
\def\IF{\rlx\hbox{\rm I\kern-.18em F}}
\def\IG{\rlx\hbox{\,$\inbar\kern-.3em{\rm G}$}}
\def\IH{\rlx\hbox{\rm I\kern-.18em H}}
\def\II{\rlx\hbox{\rm I\kern-.18em I}}
\def\IK{\rlx\hbox{\rm I\kern-.18em K}}
\def\IL{\rlx\hbox{\rm I\kern-.18em L}}
\def\one{\hbox{{1}\kern-.25em\hbox{l}}}
\def\0#1{\relax\ifmmode\mathaccent"7017{#1}%
B        \else\accent23#1\relax\fi}

                \def\/{\frac}

                \def\({\Big(}
                \def\){\Big)}
                \def\[{\Big[}
                \def\]{\Big]}

  \def\rlx{\relax\leavevmode}
                \def\inbar{\vrule height1.5ex width.4pt depth0pt}
                \def\IZ{\rlx\hbox{\sf Z\kern-.4em Z}}
                \def\IR{\rlx\hbox{\rm I\kern-.18em R}}
                \def\IC{\rlx\hbox{\,$\inbar\kern-.3em{\rm C}$}}
                \def\IN{\rlx\hbox{\rm I\kern-.18em N}}
                \def\IO{\rlx\hbox{\,$\inbar\kern-.3em{\rm O}$}}
                \def\IP{\rlx\hbox{\rm I\kern-.18em P}}
                \def\IQ{\rlx\hbox{\,$\inbar\kern-.3em{\rm Q}$}}
                \def\IF{\rlx\hbox{\rm I\kern-.18em F}}
                \def\IG{\rlx\hbox{\,$\inbar\kern-.3em{\rm G}$}}
                \def\IH{\rlx\hbox{\rm I\kern-.18em H}}
                \def\II{\rlx\hbox{\rm I\kern-.18em I}}
                \def\IK{\rlx\hbox{\rm I\kern-.18em K}}
                \def\IL{\rlx\hbox{\rm I\kern-.18em L}}
                \def\one{\hbox{{1}\kern-.25em\hbox{l}}}
                \def\0#1{\relax\ifmmode\mathaccent"7017{#1}%
                B        \else\accent23#1\relax\fi}
                
\def\IC{\mathbb{C}}
\def\IR{\mathbb{R}}
\def\IK{\mathbb{K}}

\def\II{{\mathbb I}}

\def\sl{\mathfrak{sl}}

\let\phi=\varphi

\def\be{\begin{eqnarray}}
\def\ee{\end{eqnarray}}

\def\>{\rangle}
\def\<{\langle}

\begin{document}
\vspace{.2in}
\begin{center}
{\large\bf Some comments on the integrability of the noncommutative generalized massive Thirring model}\footnote{Poster contribution to the 5th International School on Field Theory and Gravitation,
April 20 - 24, 2009, Cuiab\'a city, Brazil}
\end{center}

\begin{center}
H. Blas

Instituto de F\'isica - Universidade Federal do Mato Grosso\\
Universidade Federal de Mato Grosso\\
Av. Fernando Correa, s/n, Coxip\'o \\
78060-900, Cuiab\'a - MT - Brazil\\
H. L. Carrion\footnote{Permanent address: Escola de Ci\^encias e Tecnologia / ECT-UFRN
Av. Hermes da Fonseca, 1111, Tirol
59014-615 Natal-RN.  } \\
        Instituto de F\'{\i}sica, Universidade de S\~ao Paulo,\\
   Caixa Postal 68528, 21941-972 S\~ao \,Paulo, Brazil\\
B. M. Cerna \\
        Departamento de Ciencias, Escuela de Matem\'aticas\\
        Universidad Nacional Santiago Antunez de Mayolo\\
        Huaraz-Per\'u\\
\end{center}

\vspace{2 cm}

\begin{abstract}

\vskip .1in
 Some properties of a non-commutative version of the
  generalized massive Thirring theory (NCGMT) are studied. We develop explicit calculations for the affine Lie algebra $gl(3)$ case. The  NCGMT
     model is written in terms of Dirac type fields
  corresponding to the Moyal product extension of the ordinary multi-field massive Thirring model. We discuss the Lagrangian formulation, its zero-curvature representation and integrability property of certain submodels.
\end{abstract}





\par \vskip .3in \noindent

\newpage

The massive Thirring (MT) model and its dual the sine-Gordon model are known to be integrable. Some versions of their multi-field extensions
 have appeared in the literature (see e.g. \cite{jhep1}). On the other hand, the noncommutative (NC) version of the MT model has been proposed recently \cite{jhep2}. Here we discuss some properties of the affine Kac-Moody algebraic formulation
of a noncommutative version of the so-called generalized massive Thirring model.

\section{A version of the noncommutative generalized massive Thirring model (NCGMT)}
\label{ncmt1sec}

The ref. \cite{jhep2} presents a version of the so-called NC massive Thirring model. Here we address the non-commutative version of the generalized model presented in \cite{jhep1}. It is a two-dimensional spinor model in which the usual Thirring self-interaction term for each species is extended with similar interactions among different massive spinor field species. The action for the NCGMT model in terms of the massive Thirring
field components becomes
\begin{eqnarray}
S_{NCMT} &=& \int dx^2 \sum_{i=1}^{i=3} \{ [2 i
\widetilde{\psi}^{i}_{L} \star \partial_{+}{\psi}^{i}_{L} +
2i \widetilde{\psi}^{i}_{R} \star
\partial_{-}{\psi}^{i}_{R} +i m_{i} (\widetilde{\psi}^{i}_{L} \star
{\psi}^{i}_{R}
- \psi^{i}_{L} \star \widetilde{\psi}^{i}_{R} )] \nonumber \\
\label{ac}
&-&  2(A_{L}^{i} \star A_{R}^{i}) \},
\end{eqnarray}
the star $\star$ represents the  Moyal product and
\begin{eqnarray}
 A_{R}^{1} &=& \sqrt[4]{\frac{\alpha_{1} \beta_{1}}{4}} \,\psi^{1}_{R} \star
\tilde{\psi}^{1}_{R} +
 \sqrt[4]{\frac{\beta_{3} \alpha_{3}}{4}} \, \psi^{3}_{R} \star
 \tilde{\psi}^{3}_{R} \\
A_{R}^{2} &=&  \sqrt[4]{\frac{\alpha_{2} \beta_{2}}{4}} \,
\psi^{2}_{R} \star \tilde{\psi}^{2}_{R} -\sqrt[4]{\frac{\alpha_{1}
\beta_{1}}{4}} \, \tilde{\psi}^{1}_{R} \star
 {\psi}^{1}_{R} \\
 A_{R}^{3} &=&\sqrt[4]{\frac{\beta_{3} \alpha_{3}}{4}}\, \tilde{\psi}^{3}_{R} \star
{\psi}^{3}_{R} +
 \sqrt[4]{\frac{\alpha_{2} \beta_{2}}{4}}\, \tilde{\psi}^{2}_{R} \star
 {\psi}^{2}_{R}.\\
 A_{L}^{1} &=& \sqrt[4]{\frac{\delta_{1} \lambda_{1}}{4}} \,\psi^{1}_{L} \star
\tilde{\psi}^{1}_{L} +
 \sqrt[4]{\frac{\delta_{3} \lambda_{3}}{4}} \, \psi^{3}_{L} \star
 \tilde{\psi}^{3}_{L} \\
A_{L}^{2} &=&  \sqrt[4]{\frac{\delta_{2} \lambda_{2}}{4}} \,
\psi^{2}_{L} \star \tilde{\psi}^{2}_{L} -\sqrt[4]{\frac{\delta_{1}
\lambda_{1}}{4}} \, \tilde{\psi}^{1}_{L} \star
 {\psi}^{1}_{L} \\
 A_{L}^{3} &=&\sqrt[4]{\frac{\delta_{3} \lambda_{3}}{4}}\, \tilde{\psi}^{3}_{L} \star
{\psi}^{3}_{L} +
 \sqrt[4]{\frac{\delta_{2} \lambda_{2}}{4}}\, \tilde{\psi}^{2}_{L} \star
 {\psi}^{2}_{L}.
\end{eqnarray}
Now we write the equations of motion
\begin{eqnarray}\label{eqs11a}
\partial_{+} \psi^{3}_{L} &=& -\frac{1}{2} m_{3} \psi^{3}_{R} -
i \sqrt[4]{\frac{\delta_{3}\lambda_{3}}{4}} \,  \{ \psi^{3}_{L}
\star A^{3}_{R} + A^{1}_{R}
\star\psi^{3}_{L} \}\\
 \label{eqs12a}
\partial_{+} \tilde{\psi}^{1}_{L} &=& -\frac{1}{2} m_{1}
\tilde{\psi}^{1}_{R} +  i
\sqrt[4]{\frac{\delta_{1}\lambda_{1}}{4}}\, \{
\tilde{\psi}^{1}_{L} \star A^{1}_{R} -
 A^{2}_{R} \star\tilde{\psi}^{1}_{L} \}\\
 \label{eqs13a}
\partial_{+} \tilde{\psi}^{2}_{L} &=& - \frac{1}{2} m_{2}
\tilde{\psi}^{2}_{R} + i
\sqrt[4]{\frac{\delta_{2}\lambda_{2}}{4}}\, \{
\tilde{\psi}^{2}_{L} \star A^{2}_{R} + A^{3}_{R}
\star\tilde{\psi}^{2}_{L} \}.\\
\label{eqs11b}
\partial_{+} \tilde{\psi}^{3}_{L} &=& -\frac{1}{2} m_{3} \tilde{\psi}^{3}_{R}
+ i  \sqrt[4]{\frac{\delta_{3}\lambda_{3}}{4}}\,  \{ A^{3}_{R}
\star
\tilde{\psi}^{3}_{L}  + \tilde{\psi}^{3}_{L} \star A^{1}_{R}  \}\\
 \label{eqs12b}
\partial_{+} {\psi}^{1}_{L} &=& -\frac{1}{2} m_{1}
{\psi}^{1}_{R} -  i \sqrt[4]{\frac{\delta_{1}\lambda_{1}}{4}}\, \{
A^{1}_{R} \star {\psi}^{1}_{L}  - {\psi}^{1}_{L} \star A^{2}_{R}
 \}\\
 \label{eqs13b}
\partial_{+} {\psi}^{2}_{L} &=& - \frac{1}{2} m_{2}
{\psi}^{2}_{R} - i \sqrt[4]{\frac{\delta_{2}\lambda_{2}}{4}} \{
A^{2}_{R} \star {\psi}^{2}_{L} + {\psi}^{2}_{L} \star A^{3}_{R}
\}.\\
\label{eq21a}
\partial_{-} \psi^{3}_{R} &=& \frac{1}{2} m_{3} \psi^{3}_{L} - i
\sqrt[4]{\frac{\alpha_{3}\beta_{3}}{4}} \,
 \{ \psi^{3}_{R} \star A^{3}_{L} + A^{1}_{L} \star\psi^{3}_{R} \}
\\\label{eq22a}
\partial_{-} \tilde{\psi}^{1}_{R} &=&  \frac{1}{2} m_{1}
\tilde{\psi}^{1}_{L} + i\sqrt[4]{\frac{\alpha_{1}\beta_{1}}{4}}
 \, \{ \tilde{\psi}^{1}_{R} \star A^{1}_{L}
 - A^{2}_{L} \star\tilde{\psi}^{1}_{R}
\}\\
\label{eq23a}
\partial_{-} \tilde{\psi}^{2}_{R} &=& \frac{1}{2} m_{2}
\tilde{\psi}^{2}_{L}
 + i \sqrt[4]{\frac{\alpha_{2}\beta_{2}}{4}}
 \, \{\tilde{\psi}^{2}_{R} \star A^{2}_{L}
 + A^{3}_{L} \star\tilde{\psi}^{2}_{R} \}.
\\
\label{eq21b}
\partial_{-} \tilde{\psi}^{3}_{R} &=& \frac{1}{2} m_{3} \tilde{\psi}^{3}_{L} + i
\sqrt[4]{\frac{\alpha_{3}\beta_{3}}{4}}\,
 \{ A^{3}_{L} \star
\tilde{\psi}^{3}_{R}  + \tilde{\psi}^{3}_{R} \star A^{1}_{L} \}
\\\label{eq22b}
\partial_{-} {\psi}^{1}_{R} &=&  \frac{1}{2} m_{1}
{\psi}^{1}_{L} - i \sqrt[4]{\frac{\alpha_{1}\beta_{1}}{4}}
 \, \{ A^{1}_{L} \star
{\psi}^{1}_{R}  - {\psi}^{1}_{R} \star A^{2}_{L}
\}\\
\label{eq23b}
\partial_{-} {\psi}^{2}_{R} &=& \frac{1}{2} m_{2}
{\psi}^{2}_{L}
 - i \sqrt[4]{\frac{\alpha_{2}\beta_{2}}{4}}
 \, \{A^{2}_{L} \star
{\psi}^{2}_{R}  + {\psi}^{2}_{R} \star A^{3}_{L} \}.
\end{eqnarray}
The set of  equations of motion above are the  $gl(3)$ extension
of the equations of motion given before for the case  $gl(2)$ NCMT$_{1}$ ( see eqs.
(5.11)-(5.14) of ref. \cite{jhep2}). In fact, the later system is contained in the $gl(3)$ extended model. For example, if one considers
$\psi^{1}_{L}=\psi^{2}_{L}=\tilde{\psi}^{1}_{L}=\tilde{\psi}^{1}_{L}=0$ in the eq. (\ref{eqs11b})
then it is reproduced the equation (5.13) of reference
\cite{jhep2} describing the single Thirring field
$\psi_{3}$ provided that the parameters expression
$\sqrt[4]{\frac{\delta_{3}\lambda_{3}\beta_{3}\alpha_{3}}{16}} $
 corresponds to the coupling constant $\frac{\lambda}{2}$ of NCMT model as defined in that reference.

The four field interaction terms in the action (\ref{ac}) can be re-written as a sum of Dirac type current-current terms for the various flavors $(j=1,2,3)$. In the constructions of the relevant currents the double-gauging of a $U(1)$ symmetry in the star-localized Noether procedure deserve a
careful treatment \cite{jhep2}. So, one has two types of currents for each flavor \cite{jhep2}
\begin{eqnarray} j_{k}^{(1)\,\mu}&=&\bar{\psi}_{k}
\gamma^{\mu} \star \psi_{k}, \label{j1}\\
j_{k}^{(2)\,\mu}&=&-\psi_{k}^{T}\gamma^{0} \gamma^{\mu} \star \widetilde{\psi}_{k},\,\,\,\,\,\,\, k=1,2,3.\label{j2}.\end{eqnarray}

In order to write as a sum of current-current interaction terms it is
necessary to impose the next constraints on the coupling parameters $\frac{\delta_{j}\lambda_{j}}{\alpha_{j}\beta_{j}}= \kappa = \mbox{const.};\,\,\,\,\,\,j=1,2,3.\,\,\,\mbox{with}\,\,\,\kappa^3 =1.$
Then the current-current terms can be written as
\begin{eqnarray} \label{ncc}
-2 \sum^{3}_{i=1} A_{L}^{i} \star A_{R}^{i} &=& -g_{11} \, (j_{1\,\mu}^{(1)} \star j_{1}^{(1)\mu} +
j_{1\,\mu}^{(2)} \star j_{1}^{(2)\mu}) - g_{22} \,(j_{2\,\mu}^{(1)} \star
j_{2}^{(1)\mu} + j_{2\,\mu}^{(2)} \star j_{2}^{(2)\mu})  -\nonumber\\&&
g_{33}\,(j_{3\,\mu}^{(1} \star j_{3}^{(1)\mu} + j_{3\,\mu}^{(2)} \star
j_{3}^{(2)\mu} ) + g_{12}\,(j_{1\,\mu}^{(1)} \star
j_{2}^{(2)\mu})-\nonumber\\&&g_{23}\, ( j_{2\,\mu}^{(1)} \star j_{3}^{(1)\mu}) -
g_{13}(j_{1\,\mu}^{(2)} \star
j_{3}^{(2)\mu} ) \label{jj},
\end{eqnarray}
where
\begin{eqnarray}
g_{jj} = \frac{1}{4}  \sqrt[4]{\alpha_{j}\beta_{j}
\delta_{j} \lambda_{j}},\,\,\,\, g_{jk} = \frac{1}{2}  \sqrt[4]{\alpha_{j}\beta_{j}
\delta_{k} \lambda_{k}},\, (j\neq k);\;\,\,\,\,j,k=1,2,3.
\end{eqnarray}

The two type of U(1) currents $j_{k\,\mu}^{(1)},\, j_{k\,\mu}^{(2)}$ (k=1,2,3), respectively, satisfy the equations
\begin{eqnarray}
 \partial_{+} (\tilde{\psi}^{k}_{L} \star
{\psi}^{k}_{L})  + \partial_{-}( \tilde{\psi}^{k}_{R} \star
{\psi}^{k}_{R})  =0,\,\,\,\,\,  \partial_{+} (\psi^{k}_{L} \star \tilde{\psi}^{k}_{L})  +
   \partial_{-}( \psi^{k}_{R} \star \tilde{\psi}^{k}_{R}) =0,\,\,\,\,\,k=1,2,3
   \label{currents}.
\end{eqnarray}

\section{Matrix valued fields in the action and the zero curvature condition}
\label{zerocurvature}

We propose the NCGMT action related to the fields
$W^{\pm}_{m}$ as
\begin{eqnarray} \label{ncmt1}
S[W^{\pm}_{m} ] &=& \sum^{2}_{m=1} \int dx^2 \{ \frac{1}{2} <[
E_{-3}, W^{+}_{3-m}] \star \partial_{+} W^{+}_{m}> -
\frac{1}{2}< [ E_{3},
W^{-}_{3-m}] \star \partial_{-} W^{-}_{m}> - \nonumber \\
& &  < [E_{-3},W^{+}_{m}] \star [E_{3},W^{-}_{m}]> \}
-\frac{1}{2} \sum^{2}_{m,n=1} <\hat{J}^{+}_{m} \star
\hat{J}^{-}_{n}>.
\end{eqnarray}

The action above possesses some  global symmetries and the associated matrix-valued
currents
\begin{eqnarray}
\label{curr1}
 J^{\pm}_{m} &=& \pm \frac{1}{4}  [[E_{\mp 3}, W^{\pm}_{m}],
W^{\pm}_{3-m}]_{\star}.
\end{eqnarray}
Notice that the currents $J_{m}^{\pm}$, despite the indices in their notation, has zero gradation. The fields $W^{\pm}_{m}\,\,(m=1,2)$ and the matrix element $E_{\pm 3}$ are given in the appendix \ref{matrix}, and they have gradation $\pm m$ and $\pm 3$, respectively. The hatted fields mean that the spinor fields have been re-scaled conveniently by $\hat{W}^{\pm}_{m}= L^{\pm}_{m} W^{\pm}_{m} (L^{\pm}_{m})^{-1}$, where $L^{\pm}_{m}$ are some constant matrices.

The zero-curvature condition encodes integrability even in the NC extension of integrable models (see e.g. \cite{jhep2} and references therein), as this condition allows, for example, the construction of infinite conserved charges for them. The $gl(3)$ NCGMT  model equations of motion (\ref{eqs11a})-(\ref{eq23b}) can be formulated as a zero curvature
condition by considering the following Lax pair
\begin{eqnarray} \label{laxp}
A_{-}=E_{-3} + a[E_{-3},W^{+}_{1}]_{\star} + b
[E_{-3},W^{+}_{2}]_{\star}+ g_{1}
[[E_{-3},\hat{W}^{+}_{1}],\hat{W}^{+}_{2}]_{\star} + g_{2}
[[E_{-3},\hat{W}^{+}_{2}],\hat{W}^{+}_{1}]_{\star}.\nonumber
 \\
A_{+}=-E_{+3} + b[E_{+3},W^{-}_{1}]_{\star} +a
[E_{+3},W^{-}_{2}]_{\star}+ +\widetilde{g}_{1}
[[E_{+3},\hat{W}^{-}_{1}],\hat{W}^{-}_{2}]_{\star}
+\widetilde{g}_{2}
[[E_{+3},\hat{W}^{-}_{2}],\hat{W}^{-}_{1}]_{\star}, \nonumber
\end{eqnarray}
where $a,b, g_{1}, g_{2}, \widetilde{g}_{1}, \widetilde{g}_{2}$ are some parameters to be determined below. These matrix valued fields must be replaced into the zero-curvature equation
\begin{eqnarray}\label{curvnula}
[ \partial_{+} +  A_{+}\,, \, \partial_{-} +  A_{-} ]_{\star} = 0,
\end{eqnarray}
In order to get the relevant equations of motion it is useful to take into consideration the gradation structure of the various terms. So, we can expand explicitly the terms of gradation $(-1)$ obtaining
\begin{eqnarray}\label{cc1}
[ E_{-3}, \partial_{+}{W}^{+}_{2} ]_{\star} &= & + [E_{-3},
[E_{3},{W}^{-}_{1}]]_{\star}  -(4 g_1 + 4 g_2 )(L_{2}^{+})^{-1}
[\hat{J}^{-}_{1}, [E_{-3},\hat{W}^{+}_{2}]]_{\star}L^{+}_{2}
\end{eqnarray}
plus the constraint
\begin{eqnarray}\label{cons1}
 \Big[ F_{1}^{+},  F_{2}^{-}
\Big]_{\star}=0.
\end{eqnarray}
Next, looking for the gradation $+1$  terms we arrive at the equation
\begin{eqnarray}\label{cc2}
 \Big[ E_{3}, \partial_{-} W^{-}_{2} \Big]_{\star} & = & -
[E_{3},[E_{-3},{W}^{+}_{1}]]_{\star}  -(4 g1 +4 g2)
(L_{2}^{-})^{-1} [\hat{J}^{+}_{1},
[E_{3},\hat{W}^{-}_{2}]]_{\star} L^{-}_{2},
\end{eqnarray}
plus the constraint
\begin{eqnarray}\label{cons2}
\Big[ F_{2}^{+},  F_{1}^{-} \Big]_{\star}=0.
\end{eqnarray}
Following the process we can write similar eqs. for the $\pm 2$ gradations. We conclude
that in order to obtain the equations of motion (\ref{eqs11a})-(\ref{eq23b})
 it is required the conditions  $L_{2}^{\pm} = L_{1}^{\pm}\,$ and $g_1 = g_2 = -\frac{1}{4}$, provided that the constraints (\ref{cons1}) and (\ref{cons2}) are imposed.

 Finally, for the zero gradation term there appears the following equation
\begin{eqnarray}
\partial_{+} \hat{J}_{1}^{+} + \partial_{+} \hat{J}_{2}^{+} -
\partial_{-} \hat{J}_{1}^{-} + \partial_{-} \hat{J}_{2}^{-} - a b [F_{2}^{+}
,F_{2}^{-}]  -a b [F_{1}^{+},F_{1}^{-}] +  [\hat{J}_{1}^{-}
,\hat{J}_{1}^{+}] + [\hat{J}_{1}^{-} ,\hat{J}_{2}^{+}]
&+& \nonumber\\
\Big[\hat{J}_{2}^{-} ,\hat{J}_{1}^{+}\Big] + [\hat{J}_{2}^{-}
,\hat{J}_{2}^{+}]&=& 0.\label{cons3}
\end{eqnarray}
Taking into account the conditions
$\hat{J}_{1}^{+} = \hat{J}_{2}^{+}$ \,and \,$ \hat{J}_{1}^{-} =
\hat{J}_{2}^{-}$ in the above equation we may write it as
\begin{eqnarray}\label{c3zero}
&&  \partial_{+} \hat{J}_{1}^{+} -
\partial_{-} \hat{J}_{1}^{-}  - \frac{a b}{2} [F_{2}^{+}
,F_{2}^{-}]  - \frac{a b}{2} [F_{1}^{+},F_{1}^{-}] + 2
[\hat{J}_{1}^{-} ,\hat{J}_{1}^{+}] =0.
\end{eqnarray}
The zero gradation equation must be consistent with the equations of motion described above. In order to see the form of these three equations let us write one of the them in terms of the fundamental fields
\begin{eqnarray}
 i(\partial_{+} A^{3}_{L} -
\partial_{-} A^{3}_{R} )_{\star} &=& ( A^{3}_{R} \star A^{3}_{L}- A^{3}_{L} \star A^{3}_{R})   - a b \{i m_{3} (
\sqrt[4]{\frac{\beta_{3}\lambda_{3}}{4}}\tilde{\psi}^{3}_{R} \star
{\psi}^{3}_{L}+ \sqrt[4]{\frac{\delta_{3}\alpha_{3}}{4}}
\tilde{\psi}^{3}_{L} \star {\psi}^{3}_{R}  )+\nonumber \\ && i m_{2} (
\sqrt[4]{\frac{\delta_{2}\alpha_{2}}{4}} \tilde{\psi}^{2}_{R}
\star {\psi}^{2}_{L}+ \sqrt[4]{\frac{\beta_{2}\lambda_{2}}{4}}
\tilde{\psi}^{2}_{L} \star
{\psi}^{2}_{R} ) \} \label{zero3}
\end{eqnarray}
In particular, if we reduce the eq. (\ref{zero3}) to get an equation for just one field, say $\psi^{3}$, one has
\begin{eqnarray}
- \sqrt[4]{\frac{\delta_{3}\lambda_{3}}{4}} \partial_{+}
(\tilde{\psi}^{3}_{L}\star {\psi}^{3}_{L}) +
\sqrt[4]{\frac{\beta_{3}\alpha_{3}}{4}} \partial_{-}
(\tilde{\psi}^{3}_{R}\star {\psi}^{3}_{R}) &=&  a b m_{3} (
\sqrt[4]{\frac{\beta_{3}\lambda_{3}}{4}}\tilde{\psi}^{3}_{R} \star
{\psi}^{3}_{L}+ \sqrt[4]{\frac{\delta_{3}\alpha_{3}}{4}}
\tilde{\psi}^{3}_{L} \star {\psi}^{3}_{R}  ) + \nonumber \\
&& i\sqrt[4]{\frac{\delta_{3}\alpha_{3} \beta_{3}\lambda_{3}}{16}}
(\tilde{\psi}^{3}_{R}\star {\psi}^{3}_{R}\star
\tilde{\psi}^{3}_{L}\star {\psi}^{3}_{L}-
\tilde{\psi}^{3}_{L}\star {\psi}^{3}_{L} \star
\tilde{\psi}^{3}_{R}\star {\psi}^{3}_{R}) \nonumber \\
\label{psi3eqcons}.
\end{eqnarray}
Now, taking into account the relationships $\frac{\delta_{j}\lambda_{j}}{\alpha_{j}\beta_{j}}= 1; \,\,j=1,2,3 $ with the condition $\lambda_{3} \beta_{3}=\delta_{3}\alpha_{3}$\, and  the identifications
$\sqrt[4]{\frac{\lambda_{3} \delta_{3}}{4}} \rightarrow -\lambda$, \, $ ab m_{3} \rightarrow m_{\psi}$\, we
arrive at the equation (5.18) of the ref. \cite{jhep2} .

In terms  of the fundamental fields the constraints
(\ref{cons1}) and (\ref{cons2}) have the following form
\begin{eqnarray}
\label{cons11}
\psi^{1}_{R} * \psi^{2}_{L} = \psi^{1}_{L} * \psi^{2}_{R}, \;\,\,
\psi^{2}_{R}* \tilde{\psi}^{3}_{L} = -\psi^{2}_{L} *
\tilde{\psi}^{3}_{R},\;\,\, \tilde{\psi}^{3}_{L} * \psi^{1}_{R}=-
\tilde{\psi}^{3}_{R} * \psi^{1}_{L}
\end{eqnarray}
and
\begin{eqnarray}
\psi^{3}_{R} * \tilde{\psi}^{2}_{L} = -\psi^{3}_{L} *
\tilde{\psi}^{2}_{R}, \;\,\,\tilde{\psi}^{1}_{L}* {\psi}^{3}_{R} =-
\tilde{\psi}^{1}_{R} * {\psi}^{3}_{L},\;\,\, \tilde{\psi}^{2}_{R}
* \tilde{\psi}^{1}_{L}= \tilde{\psi}^{2}_{L} * \tilde{\psi}^{1}_{R},
\label{cons22}\end{eqnarray}
respectively.

The action (\ref{ac}) (or its matrix form (\ref{ncmt1})) defines a three species NC generalized massive Thirring model since one has six complex fields, i.e. $\psi_{R,L}^{j}$ and its complex conjugates
$\widetilde{\psi}_{R,L}^{j},\,\,(j=1,2,3)$. The zero-curvature formulation requires the above six constraints (\ref{cons11}) and (\ref{cons22}). This fact suggests that the NCGMT model defined by the action (\ref{ac}) becomes integrable only for a submodel defined by the eqs. of motion (\ref{eqs11a})-(\ref{eq23b}) provided the constraints (\ref{cons1}) and (\ref{cons2}) are satisfied. So, one expects that a careful introduction of the constraints trough relevant Lagrange multipliers into the action will provide the lagrangian formulation of an integrable submodel of the NCGMT theory.

Regarding the action related  to  the zero curvature equations without constraints it is interesting to notice that the quadratic terms in the equations of
motion make it difficult to believe that one can find a local
lagrangian for the theory. Obviously, in that case we could not
have a generalized massive Thirring model with a local lagrangian
involving bilinear (kinetic and mass terms) and current-current
terms.

{\bf Acknowledgments}

HB thanks CNPq for partial support. HLC thanks  FAPESP for support during the initial stage of the work.

\appendix

\section{The sl(3) affine Kac-Moody algebra and the matrix fields}
\label{matrix}

The matrix fields entering the potentials take the form
\begin{eqnarray} F^{\pm}_{m}&=&
\mp [E_{\pm 3}\,,\, W_{3-m}^{\mp}]\label{fw},\,\,\,m=1,2\\
 E_{\pm 3}&=& \frac{1}{6} [(2 m_{1}+m_{2}) H^{\pm 1}_{1} +
(2m_{2}
+ m_{1}) H^{\pm 1}_{2} ],\,\,\,\,\,\,\,m_{3}=m_{1}+ m_{2} \label{ee1}\\
W^{-}_{1}&=& -\sqrt{\frac{4i}{m_{3}}} \psi^{3}_{R}
E^{-1}_{\alpha3} + \sqrt{\frac{4i}{m_{1}}}\widetilde{\psi}^{1}_{R}
E^{0}_{-\alpha1}+ \sqrt{\frac{4i}{m_{2}}}\widetilde{\psi}^{2}_{R}
E^{0}_{-\alpha2}
\label{ww1}\\
W^{+}_{1}&=& \sqrt{\frac{4i}{m_{1}}}\psi^{1}_{L} E^{0}_{\alpha1} +
\sqrt{\frac{4i}{m_{2}}}\psi^{2}_{L} E^{0}_{\alpha2}-
\sqrt{\frac{4i}{m_{3}}}\widetilde{\psi}^{3}_{L} E^{1}_{-\alpha3}
\label{ww2}\\
W^{-}_{2}&=& -\sqrt{\frac{4i}{m_{1}}} \psi^{1}_{R}
E^{-1}_{\alpha1} - \sqrt{\frac{4i}{m_{2}}}{\psi}^{2}_{R}
E^{-1}_{\alpha2}+ \sqrt{\frac{4i}{m_{3}}}\widetilde{\psi}^{3}_{R}
E^{0}_{-\alpha3}
\label{ww3}\\
W^{+}_{2}&=& \sqrt{\frac{4i}{m_{3}}}  \psi^{3}_{L} E^{0}_{\alpha3}
- \sqrt{\frac{4i}{m_{1}}}\widetilde{\psi}^{1}_{L}
E^{1}_{-\alpha1}- \sqrt{\frac{4i}{m_{2}}}\widetilde{\psi}^{2}_{L}
E^{1}_{-\alpha2} \label{ww4}
\end{eqnarray}

$E_{\alpha _{i}}^{n},H^{n}_{1},H^{n}_{2}$ ($i=1,2,3; \,
n=0,\pm 1$) are some generators of $sl(3)^{(1)}$. The commutation relations for an
affine Lie algebra in the Chevalley basis are \begin{eqnarray}
&&\left[ \emph{H}_a^m,\emph{H}_b^n\right] =mC\frac{2}{\alpha_{a}^2}K_{a b}\delta _{m+n,0}  \label{a7}\\
&&\left[ \emph{H}_a^m,E_{\pm \alpha}^n\right] = \pm K_{\alpha
a}E_{\pm \alpha}^{m+n}
\label{a8}\\
&&\left[ E_\alpha ^m,E_{-\alpha }^n\right] =\sum_{a=1}^rl_a^\alpha
\emph{H}_a^{m+n}+\frac 2{\alpha ^2}mC\delta _{m+n,0}  \label{a9}
\\
&&\left[ E_\alpha ^m,E_\beta ^n\right] = \varepsilon (\alpha
,\beta )E_{\alpha +\beta }^{m+n};\qquad \mbox{if }\alpha +\beta
\mbox{ is a root \qquad }  \label{a10}
\\
&&\left[ D,E_\alpha ^n\right] =nE_\alpha ^n,\qquad \left[ D,\emph{H}%
_a^n\right] =n\emph{H}_a^n.  \label{a12} \end{eqnarray}


\begin{thebibliography}{99}
\bibitem{jhep1}
  H. Blas, \emph{Higher grading conformal affine Toda theory and
(generalized) sine-Gordon/massive Thirring duality}, \emph{JHEP} {\bf 03} (055) 03. [arXiv:hep-th/0306171].
\bibitem{jhep2}
H.~Blas, H.~L.~Carrion and M.~Rojas,
  \emph{Non-commutative solitons and strong-weak duality},
  \emph{JHEP} {\bf 05} (037) 03. [arXiv:hep-th/0502051].
\bibitem{jhep4} H.~Blas and H.~L.~Carrion, \emph{Solitons, kinks and extended hadron model based on
the generalized sine-Gordon theory}, \emph{JHEP} {\bf 07} (027) 01. [arXiv:hep-th/0610107].

\end{thebibliography}
\end{document}